\documentstyle[prb,preprint,aps]{revtex}
\begin{document}
\draft
\title{Elastic properties of the vortex lattice 
for a superconducting film of finite thickness}
\author{Edson Sardella}
\address{Departamento de F\'{\i}sica,
Faculdade de Ci\^encias, Universidade Estadual Paulista \\
Caixa Postal 473, 17033-360, Bauru-SP, Brazil}
\date{\today}
\maketitle
\begin{abstract}
In this paper we investigate the elastic properties of the
vortex lattice for a superconducting 
film of finite thickness. We derive
an analytic expression for the compression modulus. The
shear modulus is evaluated numerically by using both
the Pearl interaction potential, valid in the limit of
very thin film, and a potential for films of arbitrary
thickness. A comparative study of the shear moduli is carried out.
\end{abstract}

\pacs{PACS numbers: 74.60.Ec, 74.60.Ge}

\tightenlines

The problem of a vortex emerging perpendicularly to a surface of
a superconductor was first considered by Pearl.\cite{pearl66,pearl64} 
He pointed out that
the vortex-vortex interaction potential at the surface
of superconductors follows the power law $1/r$ ($1/k$
in Fourier space) at large distances. The Pearl interaction potential
has been widely used to investigate the equilibrium\cite{olive98}
and the elastic\cite{martynovich93} properties of the vortex lattice
in superconducting films.

In this work we will show that in fact
the Pearl interaction potential is capable of
describing satisfactorily the superconducting properties
of a very thin superconductor. However, we will show
that as the thickness of
the film approaches the London penetration depth
$\lambda$, the use of this potential may underestimate
significantly the value of the shear modulus for
sufficiently low induction.

It has been shown elsewhere\cite{sardella00} that the energy of
an ensemble of interacting vortices inside a superconducting film of finite
thickness $d$ and the energy of the stray fields in the vacuum
is given by

\begin{equation}
F=\frac{\Phi_0^2}{8\pi}\,\int\,\frac{d^2k}{(2\pi)^2}\,
\frac{1}{\lambda^2\alpha^2}\left \{
d+\frac{2}{\lambda^2k\alpha\left [ \alpha+k
\coth\left ( 
\frac{\alpha d}{2\lambda}
\right )
\right ]}
\right \}
|S({\bf k})|^2
\;.\label{free_energy}
\end{equation}
where ${\bf k}=(k_x,k_y)$, $k=\sqrt{k_x^2+k_y^2}$,
$\alpha=\sqrt{k^2+\lambda^{-2}}$, and $\Phi_0$ is the quantum flux.
The structure factor is given by

\begin{equation}
S({\bf k})=\sum_{i}\,e^{i{\bf k}\cdot{\bf R}_i}\;.
\end{equation}

Here ${\bf R}_i$ is the position of the $i$th vortex line.
Note that this result should be valid for an ensemble
of distorted vortices, that is, the positions of the
vortices do not necessarily correspond to the equilibrium
configuration.
The first term inside Eq.~(\ref{free_energy}) represents
the interaction energy of the vortex lines as if
the surfaces were absent.
The second term represents the surface energy
associated to the magnetic energy of the stray field
at the vacuum. Note that for
$k$ small (large $r$), $\alpha^2\sim 1/\lambda^2$.
Thus, the surface
energy goes as $\Phi_0^2/8\pi^2r$. This is the Pearl result
for vortices emerging from a semi-infinite
isotropic superconductor.\cite{pearl66}
Another interesting particular case of Eq.~(\ref{free_energy})
is the limit of a very thin film, $d\rightarrow 0$, and $k$ small.
In this limit, from Eq.~(\ref{free_energy}) it is straightforward
to show that

\begin{equation}
F=E_0\,\int\,\frac{dk^2}{(2\pi)^2}\,\frac{2\pi d}
{k\Lambda^{-1}+k^2}|S({\bf k})|^2\;,\label{pearl}
\end{equation}
where $E_0=(\Phi_0/4\pi\lambda)^2$, $\Lambda=2\lambda^2/d$.
This is precisely the energy of an ensemble of interacting vortices 
in a very thin film
first obtained by Pearl.\cite{pearl64} However, his derivation
is supposed to be valid for any $k$. In this work we use
Eq.~(\ref{pearl}) with no restriction on $k$.

The integrand of Eq.~(\ref{free_energy}) contains two
terms. The bulk term follows the power law $1/k^2$ for
large $k$. The integrand of Eq.~(\ref{pearl}) has the same
behavior as its corresponding one.
As a result, the self-energy contributions to both
Eqs.~(\ref{free_energy}) and (\ref{pearl}) diverge.
The divergence is logarithmic. 
This is so because London theory
neglects the size of the vortex-core. To remove this
short-length scale divergence we use the
Gaussian cutoff. This regularization procedure consists in
multiplying the integrand of Eq.~(\ref{free_energy}) and
(\ref{pearl}) by a factor $e^{-2\xi^2k^2}$, where
$\xi$ is the coherence length.\cite{sardella93} 
In the first case, only the bulk term needs 
a cutoff.\cite{footnote} The London picture
is valid in the limit of low induction and very strong
type II superconductor, $\kappa=\lambda/\xi\gg 1$.

Let us suppose that initially the vortex lines are located
at the equilibrium positions ${\bf R}_i^0=(X_i^0,Y_i^0)$. Let
us denote by ${\bf u}_i\equiv{\bf u}({\bf R}_i^0)$ the
displacement of the $i$th
vortex line from its equilibrium position. The new position
of the $i$th vortex line is then given by
${\bf R}_i={\bf R}_i^0+{\bf u}_i$. The elastic energy
of the distorted vortex lattice is most
conveniently written in ${\bf k}$-space by expressing the
displacement in terms of its Fourier transform

\begin{equation}
{\bf u}_i=\int_{BZ}\,\frac{dk^2}{(2\pi)^2}\,
\frac{1}{d}{\bf u}({\bf k})
e^{{\bf k}\cdot{\bf R}_i^0}\;,
\end{equation}
where the integration is taken over the first Brillouin
zone (BZ) of the
vortex lattice. Upon taking this into
Eq.~(\ref{free_energy}) and expanding it up to second order
in ${\bf u}_i$ we obtain for the excess free energy

\begin{equation}
\delta F=\frac{1}{2}\;\int\,\frac{dk^2}{(2\pi)^2}\,
\frac{1}{d}u_\alpha({\bf k})\Phi_{\alpha\beta}({\bf k})
u_\beta(-{\bf k})\;,
\end{equation}
with $(\alpha,\beta)=(x,y)$. Here $B=\Phi_0/A$ is the
induction, $A$ is the area of a unit
cell of the vortex lattice. The factor 
$1/d=\int_{-\pi/d}^{\pi/d}\,dk_z/(2\pi)$ in the integrals
was introduced deliberately to resemble the three dimensional
corresponding expressions. The coefficients
$\Phi_{\alpha\beta}({\bf k})$,
called elasticity matrix, are real, symmetric, and periodic in
${\bf k}$-space. One has

\begin{equation}
\Phi_{\alpha\beta}({\bf k})=
\frac{B^2}{4\pi}\,\sum_{\bf Q}\,
[({\bf k}+{\bf Q})_\alpha ({\bf k}+{\bf Q})_\beta
V({\bf k}+{\bf Q})-
Q_\alpha Q_\beta V({\bf Q})]\;,\label{matrix}
\end{equation}
where ${\bf Q}$ are the reciprocal lattice vectors. Although the
last equation is valid for any type of vortex lattice, we will use
a triangular lattice. The basis vectors of the reciprocal lattice 
are given by

\begin{equation}
{\bf Q}_1=\frac{2\pi}{a}\left (
\hat{\bf x}-\frac{\hat{\bf y}}{\sqrt{3}}
\right )\;,\;\;{\bf Q}_2=\frac{4\pi}{a}
\frac{\hat{\bf y}}{\sqrt{3}}\;
\end{equation}
where $a^2=2\Phi_0/\sqrt{3}\,B$. Then, the vortex positions in
reciprocal space are given by
${\bf Q}\equiv{\bf Q}_{mn}=n{\bf Q}_1+m{\bf Q}_2$,
with $m,n$ integers.

The interaction potential $V({\bf k})$ for a film of arbitrary
thickness is

\begin{equation}
V({\bf k})=\frac{1}{\lambda^2\alpha^2}\left [
1+\frac{2}{d\lambda^2k\alpha(k+\alpha)}
\right ]\;,\label{potential_full}
\end{equation}
and in the limit of Pearl is

\begin{equation}
V({\bf k})=\frac{2}
{kd+\lambda^2k^2}\;.\label{potential_pearl}
\end{equation}

Since we have assumed that the vortex lines are straight and
parallel to each other, there will be only two elastic
constants connected to the excitation of the lattice. One is
the compression modulus and the other one is the shear modulus.
Usually, shear is softer than compression. Let us first neglect
the discreteness of the lattice by considering only
the ${\bf Q}=0$ contribution to Eq.~(\ref{matrix}). Within
this continuum approximation, only the compression modulus
is obtained. One has

\begin{equation}
c_{11}({\bf k})=\frac{B^2}{4\pi}V({\bf k})\;.
\end{equation}

Note that in the long wave length limit, $k\rightarrow 0$,
the interaction potential diverges, both for the full
expression and within
the Pearl limit [see Eqs.~(\ref{potential_full}) and
(\ref{potential_pearl})]. As a result, in this local limit
the compression modulus diverges. Thus, as the
vortex density becomes very small, the energy cost to
compress the lattice is very large.

The evaluation of the more important
shear modulus is possible only if
one goes beyond the continuum limit. Over the first
Brillouin zone the shear modulus is nearly a constant,
that is, it does not contain any significant non-locality like the
compression modulus.\cite{brandt77} Thus,
either of the following definitions is sufficient to describe
the shear deformations

\begin{equation}
c_{66}=\lim_{k\rightarrow 0}\,\frac{\Phi_{yy}(k,0)}{k^2}\;,\;\;
c_{66}=\lim_{k\rightarrow 0}\,\frac{\Phi_{xx}(0,k)}{k^2}\;.
\end{equation}

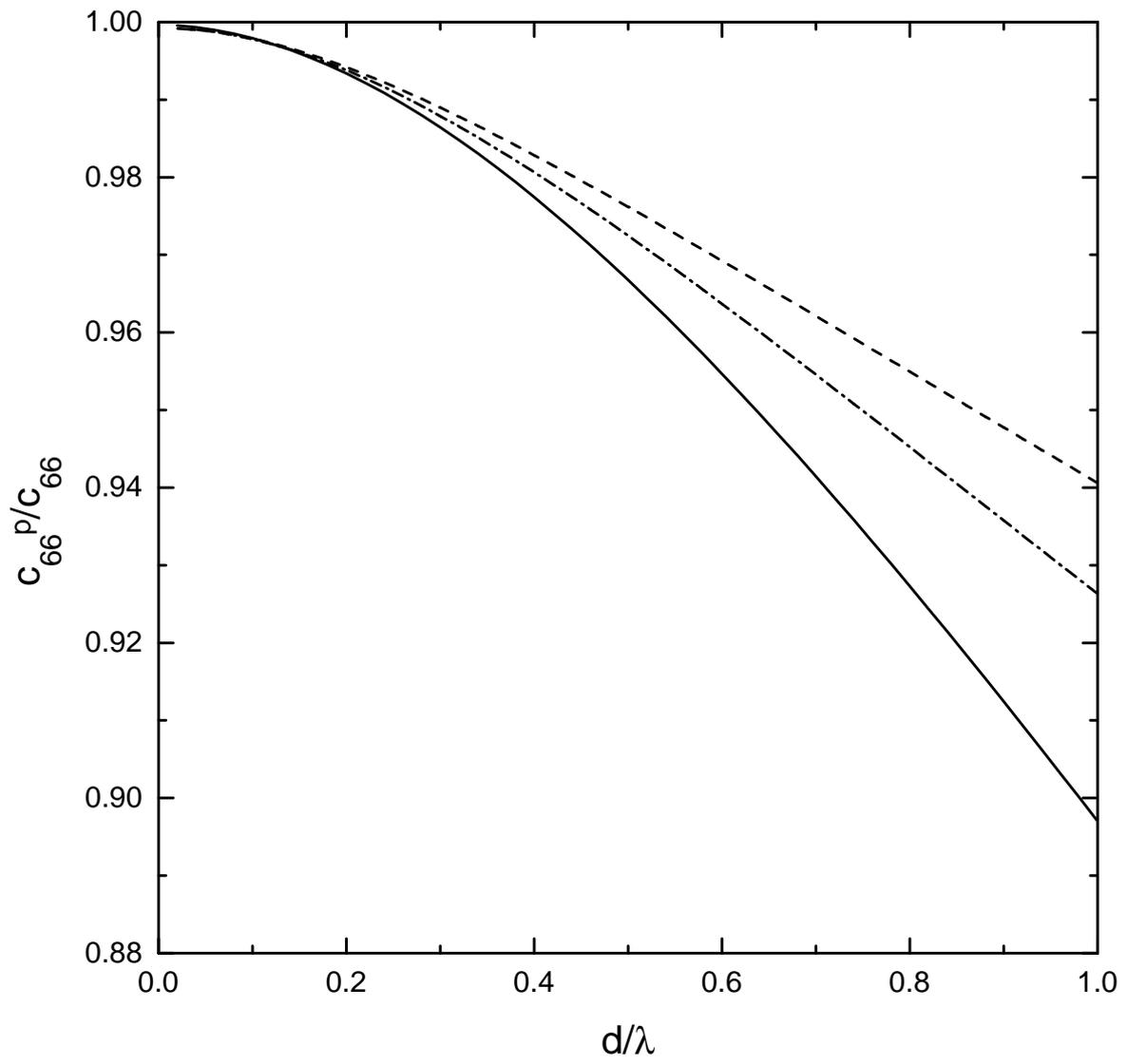
\begin{figure}
\caption{The ratio $c_{66}^p/c_{66}$ for several
values of the induction. {\it
(a)} $b=0.0005$ (continuous line); {\it (b)} $b=0.005$ 
(dot-dashed line); {\it (c)} $b=0.01$ (dashed line).} \label{fig1}
\end{figure}

In what follows, the induction $B$ is in units of the upper
critical field $H_{c2}$, $b=B/H_{c2}$; lengths are in units of
$\lambda$; the Ginzburg-Landau parameter used is $\kappa=50$. By
using these parameters we evaluated numerically $c_{66}^p$ and
$c_{66}$, where the superscript $p$ means that the Pearl
potential was used to calculate the shear modulus, whereas with no
superscript we used the full expression of
Eq.~(\ref{potential_full}). Fig.~\ref{fig1} shows the ratio
$c_{66}^{p}/c_{66}$ as a function of the film thickness. As one
can see from this figure, for sufficiently low induction the
difference between both shear modulus grows as $d$ increases. Note
that this occurs for values slightly larger than $\xi$. However,
these differences tend to disappear as the induction increases.

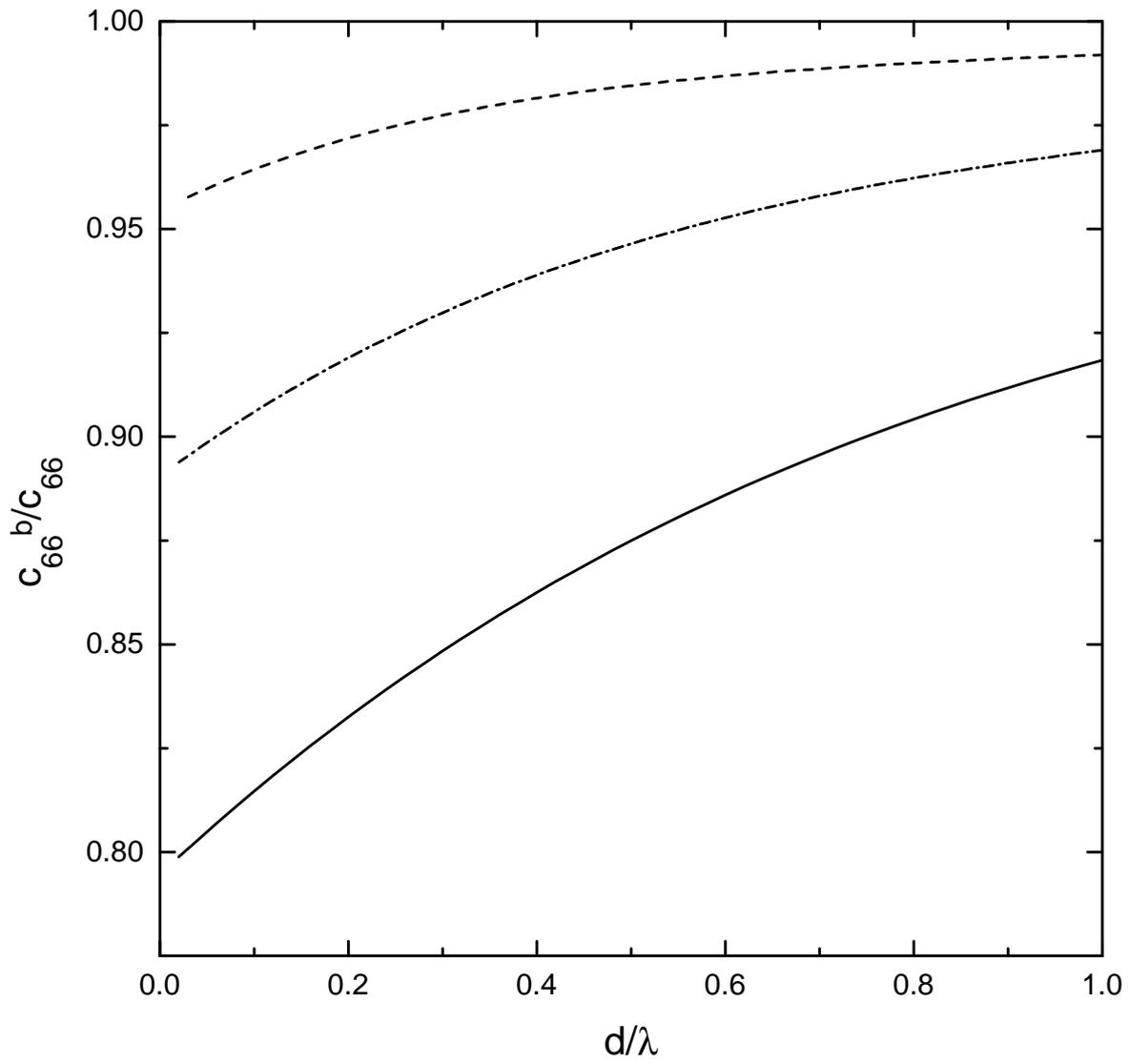
\begin{figure}
\caption{The ratio $c_{66}^b/c_{66}$ for several
values of the induction. {\it
(a)} $b=0.001$ (continuous line); {\it (b)} $b=0.002$ (dot-dashed line);
{\it (c)} $b=0.005$ (dashed line).} \label{fig2}
\end{figure}

In Fig.~\ref{fig2} we also show a plot of the ratio 
$c_{66}^b/c_{66}$ as a function of the film thickness, 
where the superscript $b$ stands for 
bulk. Note that as the induction increases the shear modulus 
the bulk shear modulus becomes approximately equal to 
full shear modulus. This was first experimentally 
observed by Fiory.\cite{fiory73}     

The conclusion we draw from this scenario
is that the Pearl potential is
indeed valid in the limit of very thin film.
However, the study of the elastic properties of the vortex lattice 
in superconducting films, of thickness slightly larger
than $\xi$ and the induction very close
to the lower critical field
$H_{c1}$, might become non-reliable by employing this
potential. In this situation it should be more
convenient to use
the full expression for the interaction potential,
but for films of thickness
not much larger than $\lambda$, because in this
limit bending of the vortex lines may become important.

\acknowledgments
The author thanks the Brazilian Agencies FAPESP and CNPq
for financial support.

\begin{references}
\bibitem{sardella00}Edson Sardella, Cond-Mat/0002152.
In this reference the physical properties are presented for a more
general situation of an anisotropic superconducting film
with the surfaces considered parallel to the $bc$ face of the
crystal and the vortices normal to it. 
By taking the isotropic limit,
$\lambda_c=\lambda_{ab}=\lambda$, in Eq.~(39) of this reference
we obtain Eq.~(\ref{free_energy}). 
Here $\lambda_{ab}$ and $\lambda_c$ are the in- and out-plane
penetration depth, respectively.
\bibitem{footnote}It is worthwhile noting that
the surface term of Eq.~(\ref{free_energy}) is
finite for a single vortex, since the
integrand goes as $1/k^5$ for large $k$. So its surface
self-energy contribution is finite.
\end{references}
\end{document}